\documentclass[%
reprint,
superscriptaddress,
 amsmath,amssymb,
 aps,
 prb,
]{revtex4-1}

\usepackage{graphicx}
\usepackage{dcolumn}
\usepackage{bm}
\usepackage{makecell}

\begin{document}

\preprint{APS/123-QED}

\title{Tunable perpendicular exchange bias in oxide heterostructures}

\author{Gideok Kim}
\affiliation{Max-Planck-Institute for Solid State Research, Heisenbergstrasse 1, 70569 Stuttgart, Germany}%
\author{Yury Khaydukov}
\affiliation{Max-Planck-Institute for Solid State Research, Heisenbergstrasse 1, 70569 Stuttgart, Germany}%
\affiliation{Max Planck Society, Outstation at the Heinz Maier-Leibnitz (MLZ), 85748 Garching, Germany}%
\author{Martin Bluschke}
\affiliation{Max-Planck-Institute for Solid State Research, Heisenbergstrasse 1, 70569 Stuttgart, Germany}%
\affiliation{Helmholtz-Zentrum Berlin für Materialien und Energie, Wilhelm-Conrad-Röntgen-Campus BESSY II, Albert-Einstein-Strasse 15, 12489 Berlin, Germany}%
\author{Y. Eren Suyolcu}
\affiliation{Max-Planck-Institute for Solid State Research, Heisenbergstrasse 1, 70569 Stuttgart, Germany}%
\author{Georg Christiani}
\affiliation{Max-Planck-Institute for Solid State Research, Heisenbergstrasse 1, 70569 Stuttgart, Germany}%
\author{Kwanghyo Son}
\affiliation{Max Planck Institute for Intelligent systems, Heisenbergstrasse 1, 70569 Stuttgart, Germany}%
\author{Christopher Dietl}
\affiliation{Max-Planck-Institute for Solid State Research, Heisenbergstrasse 1, 70569 Stuttgart, Germany}%
\author{Thomas Keller}
\affiliation{Max-Planck-Institute for Solid State Research, Heisenbergstrasse 1, 70569 Stuttgart, Germany}%
\affiliation{Max Planck Society, Outstation at the Heinz Maier-Leibnitz (MLZ), 85748 Garching, Germany}%
\author{Eugen Weschke}
\affiliation{Helmholtz-Zentrum Berlin für Materialien und Energie, Wilhelm-Conrad-Röntgen-Campus BESSY II, Albert-Einstein-Strasse 15, 12489 Berlin, Germany}%
\author{P. A. van Aken}%
\affiliation{Max-Planck-Institute for Solid State Research, Heisenbergstrasse 1, 70569 Stuttgart, Germany}%
\author{Gennady Logvenov}
\affiliation{Max-Planck-Institute for Solid State Research, Heisenbergstrasse 1, 70569 Stuttgart, Germany}%
\author{Bernhard Keimer}%
\email{B.Keimer@fkf.mpg.de}
\affiliation{Max-Planck-Institute for Solid State Research, Heisenbergstrasse 1, 70569 Stuttgart, Germany}%

\date{\today}

\begin{abstract}
The exchange bias effect is an essential component of magnetic memory and spintronic devices. Whereas recent research has shown that anisotropies perpendicular to the device plane provide superior stability against thermal noise, it has proven remarkably difficult to realize perpendicular exchange bias in thin-film structures. Here we demonstrate a strong perpendicular exchange bias effect in heterostructures of the quasi-two-dimensional canted antiferromagnet La$_2$CuO$_4$ and ferromagnetic (La,Sr)MnO$_3$ synthesized by ozone-assisted molecular beam epitaxy. The magnitude of this effect can be controlled via the doping level of the cuprate layers. Canted antiferromagnetism of layered oxides is thus a new and potentially powerful source of uniaxial anisotropy in magnetic devices. 


\end{abstract}

\maketitle


\section{Introduction}
Exchange bias arises at the interface between ferromagnets and antiferromagnets, and manifests itself as a shift of the magnetic hysteresis loop in the direction opposite to the applied cooling field. Exchange-bias structures serve diverse functions in magnetic memory and spintronic devices and are of topical interest in both fundamental and applied research \cite{Zhang2016}. In most cases, the exchange bias is observed when the field is applied parallel to the interface. However, recent research has focused on systems with a bias direction perpendicular to the interface, because they are less susceptible to thermal noise and particularly well suited for a large class of spintronic devices \cite{Zhang2016,Maat2001,Lamirand2013,Zilske2017,Radu2012,Dieny2017}. Most of these systems utilize ferromagnets with easy axes perpendicular to the interface – an uncommon situation that requires elaborate strategies to manipulate the magneto-crystalline anisotropy. Some such strategies take advantage of interfacial anisotropies in ultrathin ferromagnetic films \cite{Maat2001,Lamirand2013,Zilske2017}; others use ferrimagnets including rare-earth species with large single-ion anisotropies \cite{Radu2012}. However, a simpler and more robust strategy based on the intrinsic properties of the components is desirable to design versatile devices.
\begin{figure*}[!htb]
	\includegraphics[width=7 in]{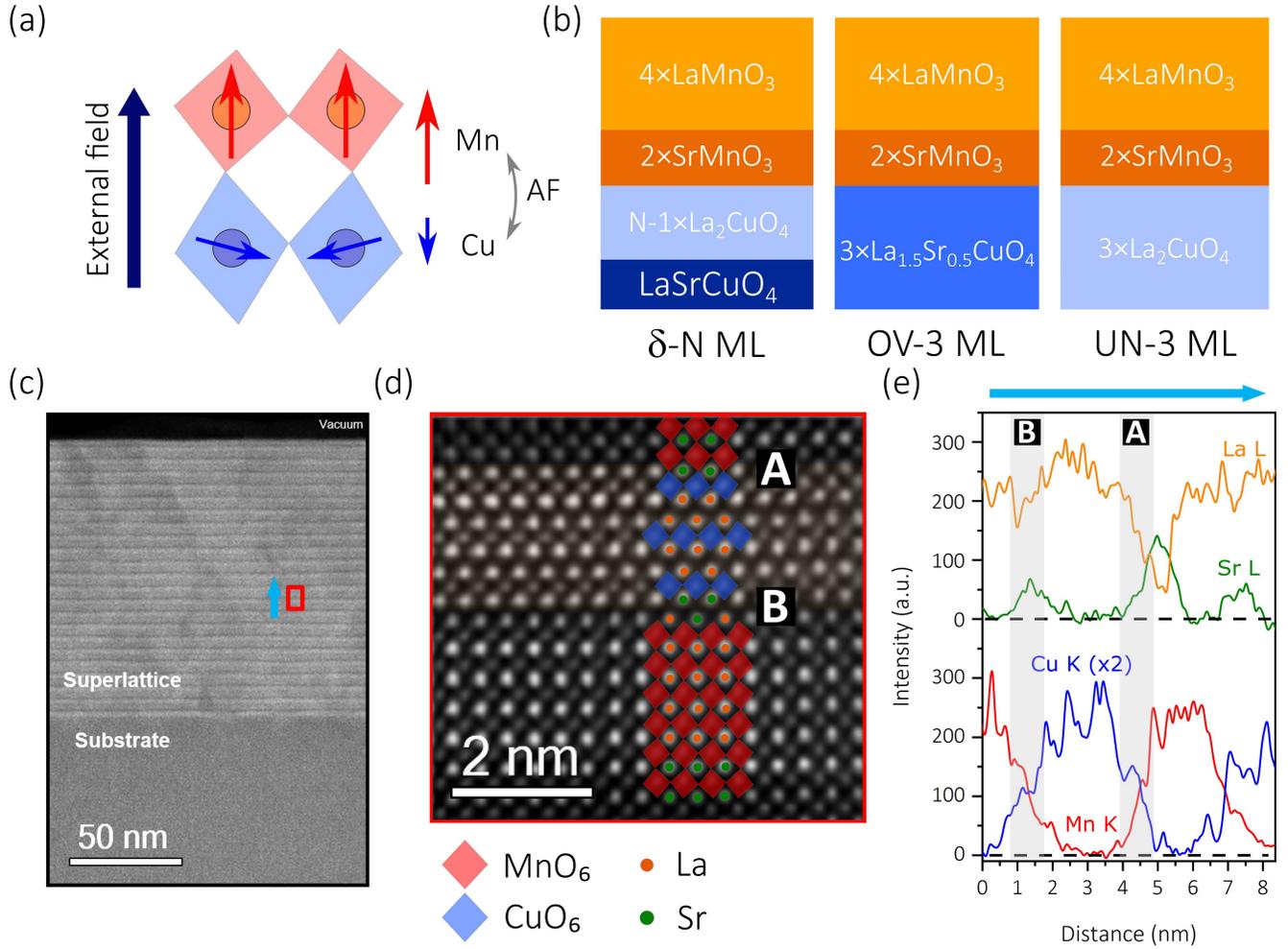}
	\caption{\label{fig:structure}Lattice structure and chemical profile of the cuprate/manganate superlattices. (a) Schematic diagram for the out-of-plane antiferromagnetic coupling between Cu and Mn. The arrows indicate the local magnetic moments at the interface (left) and the net moments in the direction of the external field (right). (b) Supercell composition of the superlattices. (c) Low magnification STEM-HAADF image of $\delta$-3 ML. The red box and the blue arrow indicate regions covered by panels (d) and (e), respectively. (d) High-magnification STEM-HAADF image of $\delta$-3 ML. (e) EDXS depth profile of $\delta$-3 ML.}
\end{figure*}
Recent advances in metal-oxide heterostructures offer new perspectives for electronic devices based on collective quantum phenomena such as unconventional magnetism, multiferroicity, and superconductivity \cite{Chakhalian2007,Yu2010,Bibes2011,Hwang2012,Bhattacharya2014}. In particular, exchange-bias structures based on ferromagnetic manganates of composition La$_{1-y}$Sr$_{y}$MnO$_3$ (0.1$\leq$$y$$\leq$0.5) and different metal-oxide antiferromagnets have been reported \cite{Wu2010,Gibert2012,Fan2016}. La$_{1-y}$Sr$_{y}$MnO$_3$ (LSMO) is a soft ferromagnet, and in thin-film form, it generally orders with magnetization direction in the substrate plane. It was recently shown, however, that perpendicular exchange bias can be induced in nanocomposite films of La$_{0.7}$Sr$_{0.3}$MnO$_3$ and antiferromagnetic LaFeO$_3$ with active interfaces perpendicular to the substrate plane \cite{Fan2016}.

\begin{figure*}[t]
	\includegraphics[width=7 in]{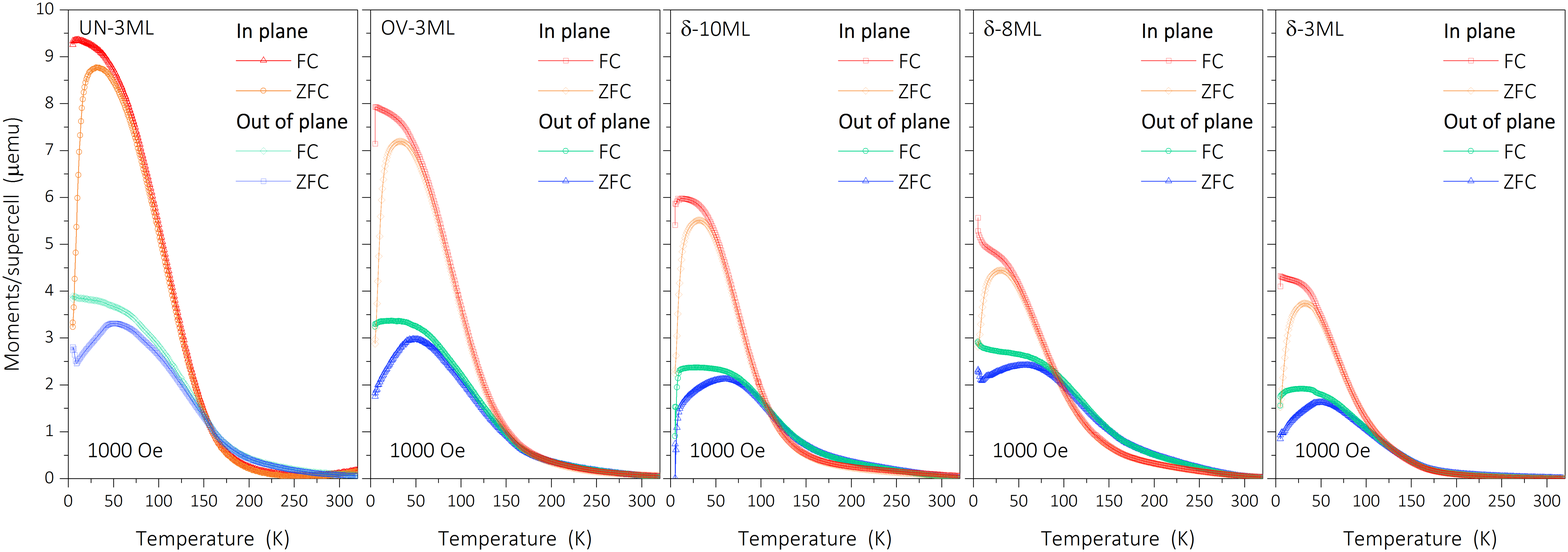}
	\caption{\label{fig:MT}Temperature dependent magnetization curves. The data were taken in field-cooled (FC) and zero-field-cooled (ZFC) modes with a 1000 Oe field applied parallel and perpendicular to the heterostructure plane, respectively. The signal from the substrate was subtracted after the measurement, and the resulting magnetic moment was normalized by the number of supercells.}
\end{figure*}

Here we report perpendicular exchange bias in a different oxide heterostructure system with a conventional layer architecture that does not require elaborate synthesis conditions. Instead of
manipulating the easy axis of the ferromagnet, the perpendicular anisotropy in our system is generated by canted moments in the quasi-two-dimensional antiferromagnet La$_2$CuO$_4$ (LCO) that are exchange-coupled to the ferromagnetic magnetization of La$_{1-y}$Sr$_{y}$MnO$_3$ (Figure \ref{fig:structure}(a)). We also show that the magnitude of the exchange bias can be tuned via the doping level of LCO.

\section{Experimental details}
Superlattices were grown on LaSrAlO$_4$ (LSAO) (001) single-crystalline substrates (Crystal GmbH) by using the ozone-assisted ALL-MBE system (DCA Instruments). The growths were monitored by using \textit{in situ} reflection high energy electron diffraction (RHEED). The film quality was confirmed by high-resolution x-ray diffraction and transmission electron microscopy.     
The full compositions of the samples are shown in table I.
The total film thickness varies among the samples, however its impact on the interfacial exchange interaction that induces the exchange-bias observed in our study is insignificant.  

\begin{table*}[!htb]
\caption{\label{tab:table1}%
List of superlattices. 
The number of layers is counted in units of one CuO$_2$ layer in La$_{2-x}$Sr$_{x}$CuO$_4$ (LSCO) ($\sim$ 6.6 \AA) and pseudo-cubic unitcell of manganates (One manganese atom per cell, $\sim$ 3.9 \AA).  
}
\begin{ruledtabular}
\begin{tabular}{ccc}
\textrm{Sample}&
\textrm{Composition}&
\textrm{Substrate}\\
\colrule
$\delta$-2ML & \makecell{9$\times$[1$\times$LSCO(x=1)+1$\times$LCO+2$\times$SMO+4$\times$LMO]}& LSAO (001) \\
$\delta$-3ML & \makecell{22$\times$[1$\times$LSCO(x=1)+2$\times$LCO+2$\times$SMO+4$\times$LMO]}& LSAO (001) \\
$\delta$-5ML & \makecell{10$\times$[1$\times$LSCO(x=1)+4$\times$LCO+2$\times$SMO+4$\times$LMO]}& LSAO (001) \\
$\delta$-8ML & \makecell{22$\times$[1$\times$LSCO(x=1)+7$\times$LCO+2$\times$SMO+4$\times$LMO]}& LSAO (001) \\
$\delta$-10ML & \makecell{22$\times$[1$\times$LSCO(x=1)+9$\times$LCO+2$\times$SMO+4$\times$LMO]}& LSAO (001) \\
UN-3ML & \makecell{9$\times$[3$\times$LCO+2$\times$SMO+4$\times$LMO]}& LSAO (001) \\
OV-3ML & \makecell{9$\times$[3$\times$LSCO(x=0.5)+2$\times$SMO+4$\times$LMO]}& LSAO (001) \\
LMO/SMO & \makecell{10$\times$[2$\times$SMO+4$\times$LMO]}& STO (001) \\

\end{tabular}
\end{ruledtabular}
\end{table*}

For scanning transmission electron microscopy (STEM), we prepared representative cross-sectional electron transparent specimens by employing the standard specimen preparation procedure including mechanical grinding, tripod wedge polishing, and argon ion milling. 
After the specimens were thinned down to $\sim$10 $\mu$m by tripod polishing, argon ion beam milling, for which a precision ion polishing system (PIPS II, Model 695) was used at low temperature, was carried out until reaching electron transparency. 
For all STEM analyses, a probe-aberration-corrected JEOL JEM-ARM200F equipped with a cold field-emission electron source, a probe Cs-corrector (DCOR, CEOS GmbH), a large solid-angle JEOL Centurio SDD-type energy-dispersive X-ray spectroscopy (EDXS) detector was used. 
STEM image and EDXS analyses were performed at probe semi-convergence angles of 20 mrad and 28 mrad, resulting in probe sizes of 0.8 $\AA$ and 1.0 $\AA$, respectively. 
The collection angle range for high-angle annular dark-field (HAADF) images was 75-310 mrad and in order to decrease the noise level, the images were processed with a principal component analysis routine.

We utilized SQUID magnetometry, polarized neutron reflectometry (PNR) and X-ray magnetic circular dichroism (XMCD) for magnetic property measurements. 
The magnetization curves were measured using a MPMS3 magnetometer (Quantum Design Co.) with VSM mode.
The PNR experiments were conducted at the angle-dispersive reflectometer NREX (neutron wavelength 0.428 nm) at FRM-II, Garching, Germany. 
An external magnetic field was applied parallel to the sample surface, normal to the scattering plane.
XMCD experiments were performed at the BESSY II
undulator beamline UE46-PGM1. 
The spectra were collected using both total-electron-yield and fluorescence-yield modes simultaneously. The XMCD signal is defined as (I$_+$-I$_-$)/( I$_+$+I$_-$).

\section{Results and discussion}
Cuprate-manganate superlattices have been extensively investigated as a platform for interfacial reconstructions and proximity effects, for the interplay between ferromagnetism and unconventional superconductivity, and for superconducting spintronics \cite{Bibes2011}. 
To study the exchange-bias effect, we chose superlattices based on LCO and LaMnO$_3$ (LMO) because their doping levels can be accurately controlled, and because they are well suited for epitaxial integration. 
We used ozone-assisted layer-by-layer molecular beam epitaxy to deposit a series of Sr-doped LCO-LMO superlattices with a heterogeneous doping technique.
All superlattices were prepared with identical ferromagnetic layers, 2$\times$SrMnO$_3$ + 4$\times$LaMnO$_3$ to reduce the Sr redistribution into the cuprate layers, and LCO layers with various densities of mobile holes, as summarized in Figure \ref{fig:structure}(b) \cite{Bhattacharya2008}.
In “$\delta$-doped” samples ($\delta$-N ML in Figure \ref{fig:structure}(b)), individual monolayers of highly overdoped LSCO supply holes to N monolayers of undoped LCO \cite{Baiutti2015}. 
Because of the chemical-potential difference between cuprates and manganates, interfacial charge transfer reduces the hole content in the cuprate layers such that the average doping level of these samples is in the “underdoped” regime close to the insulating antiferromagnet LCO, where superconductivity is absent or severely degraded. Indeed, mutual inductance measurements on $\delta$-N ML samples show no sign of a superconducting transition (although signatures of filamentary superconductivity with T$_c$ $\sim$20 K were observed in resistivity measurements). 
For comparison, we also synthesized superlattices based on three consecutive monolayers of undoped LCO (UN-3ML) and highly overdoped, non-superconducting La$_{1.5}$Sr$_{0.5}$CuO$_4$ (OV-3ML), respectively.


\begin{figure}[b]
	\includegraphics[width=3.4 in]{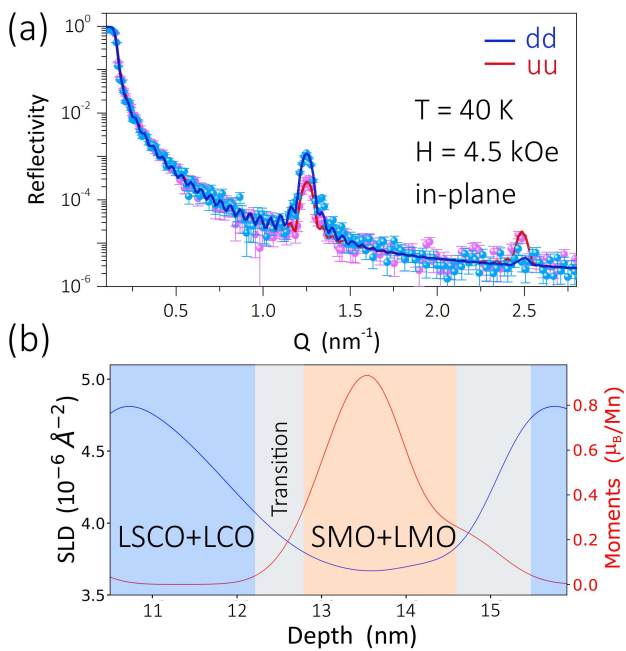}
	\caption{\label{fig:PNR}Polarized neutron reflectometry on the $\delta$-3 ML sample. (a) Polarized neutron reflectivity curves measured with up-spin, \textit{u}, and down-spin, \textit{d}, neutrons at 40 K. (b) Nuclear (left axis) and magnetic (right axis) SLD depth profiles.}
\end{figure}

\begin{figure*}[!ht]
	\includegraphics[width=7 in]{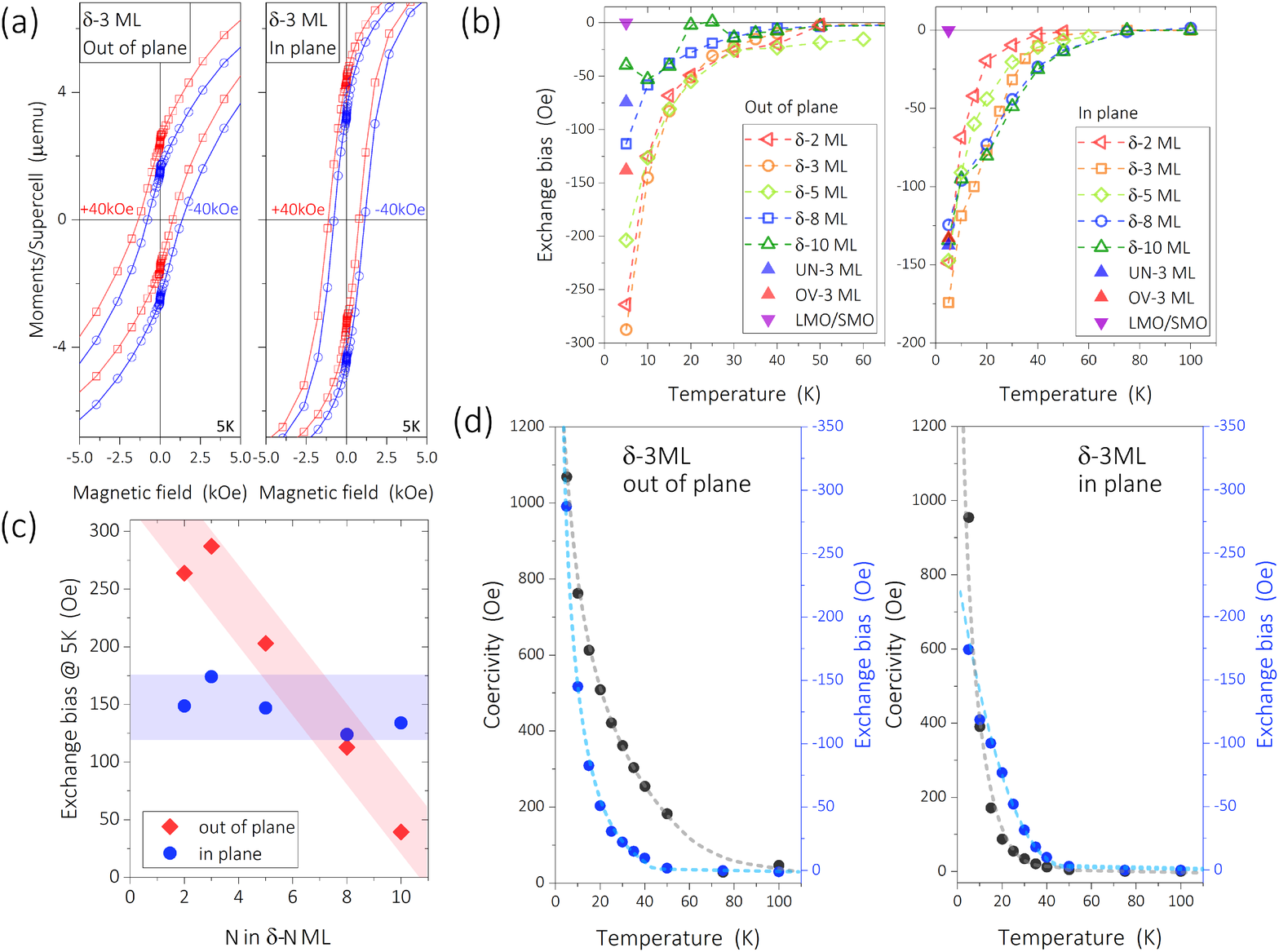}
	\caption{\label{fig:SQUID}Exchange bias and coercivity. (a) Hysteresis loops measured at 5 K showing the exchange bias effect in the $\delta$-3ML superlattice. Full curves are presented in the supplementary information. (b) Temperature dependence of H$_{EB}$ for fields applied out of (left panel) and in (right panel) the heterostructure plane. (c) Dependence of H$_{EB}$ on the thickness of the $\delta$-doped cuprate layers. (d) Temperature dependence of H$_{EB}$ and H$_{C}$ in the $\delta$-3ML sample. The external field was applied in (100)$_{LSAO}$ and (001)$_{LSAO}$  direction for in-plane and out-of-plane measurements respectively.}
\end{figure*}

Scanning transmission electron microscope high-angle annular dark-field (HAADF) images show alternating K$_2$NiF$_4$-type and perovskite structures with the intended periodicity (Figure \ref{fig:structure}(c), (d)). We observed two types of interfaces: interface A with direct Cu-O-Mn bonding that is responsible for the interfacial magnetic interaction between Cu and Mn moments, and interface B with an extra (La,Sr)-O layer that mediates the charge redistribution \cite{Wrobel2018,Uribel2014,Chakhalian2006}. The STEM energy-dispersive x-ray spectroscopy line scans show short-ranged intermixing between copper and manganese at the interfaces, which extends over less than 1 nm (Figure \ref{fig:structure}(e)).

The onset of the ferromagnetic transition in magnetization measurements revealed Curie temperatures of $\sim$170 K for all samples, consistent with prior work on SrMnO$_3$/LaMnO$_3$ superlattices (Figure \ref{fig:MT}) \cite{May2008}.
Interestingly the saturation magnetization per ferromagnetic layer, 2$\times$SrMnO$_3$ + 4$\times$LaMnO$_3$, varies with different types of cuprate spacers; the structure with LCO layers shows the largest magnetic moments, and chemical substitution in the LCO layers reduces the magnetization.
The depth-resolved profile of the in-plane magnetic moment obtained from PNR agrees with the low magnetization in the $\delta$-3ML sample, where the magnetic moment reaches up to 0.8 $\mu_B$, that is, less than half of the value $\sim$2 $\mu_B$ in optimally doped LSMO thin films (Figure \ref{fig:PNR}) \cite{Kim2012,Yuan2015}. 
The origin of the reduced magnetism can be attributed to an interface effect, because the nominal compositions of the ferromagnetic layers are identical. The largely suppressed magnetization at the interface also supports the interface-derived nature of the effect (Fig. \ref{fig:PNR}).
At the interface, both epitaxial strain and charge transfer can influence the magnetic moment in LSMO  \cite{Tsui2000,Yang2012}. 
In our case, epitaxial strain cannot be the major factor because the in plane lattice parameters of the LCO-based spacers are similar especially among samples with the same spacer thicknesses, namely $\delta$-3ML, UN-3ML and OV-3ML.
On the other hand, the strong dependence on the effective doping level of the cuprate layers indicates that charge transfer plays the major role, where holes move from the cuprate to the manganate layers to match the chemical potential difference and reduce the magnetic moments \cite{Seo2019}.


Magnetic hysteresis loops were measured after field cooling in a static magnetic field of 40 kOe. Representative curves from sample $\delta$-3ML clearly exhibit the characteristic exchange-bias shift along the magnetic field axis for both in-plane and out-of-plane applied fields (Figure \ref{fig:SQUID}(a)). To quantify the effect, the size of the exchange bias (H$_{EB}$) and the coercivity (H$_{C}$) are defined following the convention H$_{EB}$ = (H$_{C+}$ + H$_{C-}$)/2 and H$_C$ = (H$_{C+}$ - H$_{C-}$)/2, where H$_{C+}$ and H$_{C-}$ are defined as the positive and the negative H-intercepts of the M-H hysteresis loop, respectively. We found that all LCO-LMO superlattices exhibit nonzero values of H$_{EB}$ at 5K, and that H$_{EB}$ shows a strong temperature dependence that sets on below 50 K (Figure \ref{fig:SQUID}(b)). Notably, a superlattice without the cuprate (LMO/SMO) displays no signature of exchange bias, highlighting the crucial role of the interface between the cuprate and the manganite layers \footnote{Additional evidence for the presence and robustness of exchange bias in our superlattices is presented in the supplementary materials: hysteresis curves with a wider range (Fig. S1), the presence of exchange bias after several cycles of magnetization (Fig. S2), and exchange-biased hysteresis loop measured with PNR which is less susceptible to extrinsic effects than SQUID-VSM (Fig. S3)}.



We now focus on the difference between the evolution of H$_{EB}$ for in-plane and out-of-plane directions (Figure \ref{fig:SQUID}(c)). The out-of-plane exchange bias, H$_{EB,OP}$, displays a strong dependence on composition, whereas the in-plane effect, H$_{EB,IP}$, shows at most a weak composition dependence. $\delta$-N ML samples exhibit substantial anisotropies, with H$_{EB,OP}$ $>$ H$_{EB,IP}$. The anisotropy decreases continuously with increasing N (and hence decreasing doping level). Both the UN-3ML and the OV-3ML samples exhibit only small anisotropies. These findings suggest that the origins of the out-of-plane and in-plane exchange bias effects are distinct, and that the doping level selectively influences the effect along the surface normal direction. 

Figure \ref{fig:SQUID}(d) demonstrates a related anisotropy in the coercivity, H$_{C}$, which reflects the strength of the magnetic domain-wall pinning. For in-plane magnetic fields, H$_{C,IP}$ increases markedly upon cooling below 50 K, parallel to the onset of H$_{EB,IP}$, which is consistent with common EB systems \cite{Phan2016}. In contrast, H$_{C,OP}$ begins to increase at much higher temperatures (T $>$ 100 K), indicating an additional pinning mechanism. The unexpected pinning in the out of plane direction could also be inferred from the temperature dependent magnetization curves (Figure \ref{fig:MT}), where the FC curves bifurcate from the ZFC curves at higher temperatures in OP than IP suggesting enhancement of coercive field and magnetic frustration. We could find the origin of the unexpected enhancement of H$_{C,OP}$ at higher temperatures from the magnetic coupling of the ferromagnetic layer to antiferromagnetic interface layer \cite{Scholten2005}, which was reported by prior studies on cuprate/manganate heterostructures \cite{Chakhalian2006,Uribel2014,DeLuca2014}.

\begin{figure}[h]
	\includegraphics[width=3.4 in]{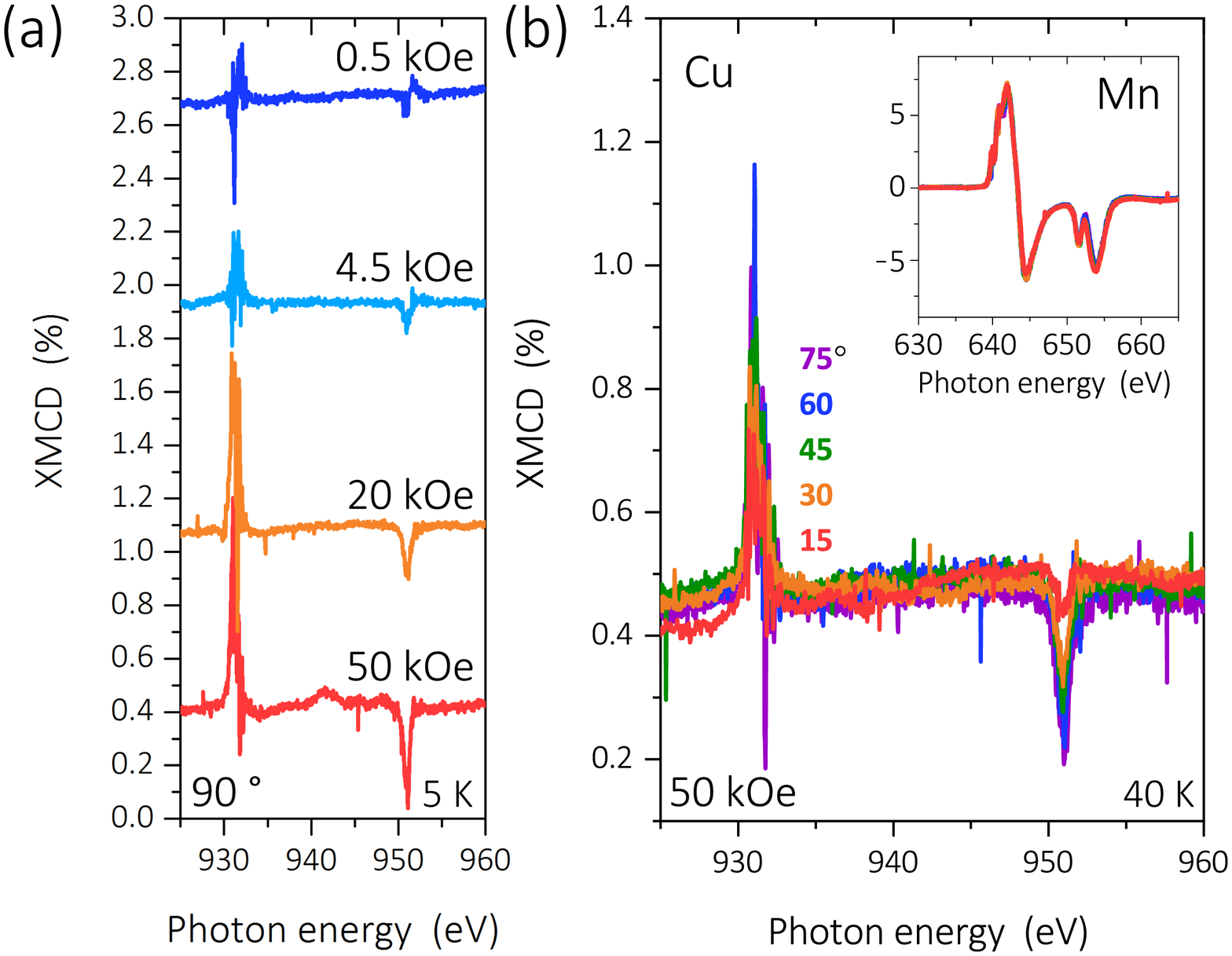}
	\caption{\label{fig:XMCD} XMCD measurements on the $\delta$-3ML superlattice in total-electron-yield mode. The current to the ground was measured as a function of incident photon energy across the Cu and Mn L$_{3,2}$ absorption edges for left- and right-circularly polarized x-rays. The magnetization of the probed sublattice is proportional to the amplitude of the XMCD signal. (a) Magnetic field dependence of the Cu-XMCD at 5 K. The magnetic field was applied perpendicular to the substrate plane. (b) Cu-XMCD spectra at 40 K in a magnetic field of 50 kOe applied at different angles to the substrate plane (see the legend). The inset shows Mn-XMCD spectra that are angle-independent.}
\end{figure}

Since neither bulk LSMO nor thin-film structures composed solely of manganates exhibit the exchange bias effect, it must be ascribed to the interaction between Mn and Cu magnetic moments across the interface (Figure \ref{fig:structure}(a)). Interfacial exchange interactions in cuprate-manganate heterostructures have been the subject of prior work with XMCD, an element-specific probe that addresses Mn and Cu moments separately \cite{Uribel2014,Chakhalian2006}. Prior XMCD measurements revealed a polarization of the Cu spins, which sets on gradually below the Curie temperature because the antiferromagnetic Cu-Mn interaction is weaker than the ferromagnetic Mn-Mn coupling. Following our observation of a highly anisotropic exchange bias and coercivity, we have carried out Cu- and Mn-XMCD experiments on the $\delta$-3ML sample that shows the largest anisotropy (Figure \ref{fig:XMCD}). For magnetic field applied perpendicular to the substrate, the Cu-XMCD data show that the Cu spins are oriented opposite to the field for low H and switch to a parallel orientation for H $<$ 4.5 kOe (Figure \ref{fig:XMCD}(a)), reflecting antiferromagnetic Cu-Mn interactions of moderate strength in agreement with prior work \cite{Uribel2014,Chakhalian2006,DeLuca2014}. 
Experiments in which a magnetic field of 50 kOe was applied in different directions with respect to the substrate plane demonstrate the out-of-plane character of the magnetic moments at the Cu site (Fig. \ref{fig:XMCD}(b)). 
The Mn-XMCD spectra in the inset of Fig. \ref{fig:XMCD}(b) show that the Mn magnetic moments at H = 50 kOe are at most weakly dependent on the magnetic field direction, as expected in view of the weak spin-space anisotropy of LMO.

`These findings suggest that the large perpendicular exchange bias and the enhanced out-of-plane coercivity originate in the perpendicular magnetization of the cuprate layers, which is exchange-coupled to the ferromagnetic magnetization of the manganate layers via interfacial interactions. A canted magnetization has indeed been observed in the CuO$_2$ sheets of antiferromagnetic bulk LCO \cite{Keimer1992,Reehuis2006,Kastner1988}. This effect arises from a cooperative tilt rotation of the CuO$_6$ octahedra in the crystal structure, which creates an inversion-asymmetric Cu-O-Cu exchange bond and actuates a Dzyaloshinskii-Moriya interaction between the Cu spins. The canted antiferromagnetism in LCO is highly sensitive to the dopant concentration. In undoped LCO, the direction of the canted moment alternates between CuO$_2$ layers, and a net canted magnetization perpendicular to the antiferromagnetic layers only appears at a metamagnetic transition in magnetic fields exceeding 10 T (100 kOe) \cite{Reehuis2006}. The high magnetic field required to reverse the direction of the canted moment is thus presumably responsible for the weak perpendicular exchange bias in the UN-3ML superlattice. The modest magnetic fields used for our experiments were not sufficiently strong to reverse canted moment directions during field cooling, so as to produce interfacial spin arrangements favorable for the perpendicular exchange bias. In doped LCO, hole doping induces a spin glass phase with short-range antiferromagnetic order and rapidly reduces the magnetic field scale for the metamagnetic transition, allowing our field cooling procedures to effectively exchange-bias the magnetic layers. In overdoped bulk LCO, the magnetic short-range order and the canted magnetization vanish entirely \cite{Keimer1992}. The onset temperature of spin-glass correlations is comparable to
the onset of the perpendicular exchange bias in our superlattices (50 K). The anomalous exchange bias can thus be attributed to interfacial moments in the spin glass, in analogy to previously reported exchange bias effects in bilayers and core/shell nanoparticles composed of ferromagnets and spin glasses \cite{Phan2016,Ali2007}. The phase behavior of bulk LCO thus provides a natural explanation for the maximal H$_{EB,OP}$ in the $\delta$-3 ML sample with underdoped LCO layers, the decrease of the anomalous perpendicular exchange bias with increasing doping (Fig. \ref{fig:SQUID}(c)), and its absence in the OV-3 ML sample.

\section{Conclusion}
In conclusion, we have shown that quasi-two-dimensional canted antiferromagnetism is a potent source of perpendicular exchange bias in metal-oxide heterostructures. 
The Dzyaloshinskii-Moriya interaction responsible for the perpendicular magnetic moments is rooted in the bulk crystal structure and is therefore more robust than magnetic anisotropies generated solely by the interfacial inversion asymmetry.
Quasi-two-dimensional antiferromagnets are quite common in metal-oxides and can be readily integrated into conventional multilayer structures, without the need to create elaborate composite architectures. 
Finally, we have shown that the magnitude of the perpendicular exchange bias can be systematically tuned by adjusting the doping level of the antiferromagnet through an atomically engineered $\delta$-doping scheme. 
Canted antiferromagnetism of layered oxides is thus a new and potentially powerful source of uniaxial anisotropy, and opens up new perspectives for spin-electronic devices that take advantage of collective quantum phenomena such as superconductivity and multiferroicity.

\begin{acknowledgments}
	We thank Peter Specht for technical support, Ute Salzberger for TEM specimen preparation and Eberhard Goering for the SQUID-VSM instrument. We acknowledge financial support by the DFG under grant TRR80.
\end{acknowledgments}

\bibliography{apssamp}

\end{document}